\documentclass[%
 reprint,
 amsmath,amssymb,
 aps,
prstper,
]{revtex4-2}

\usepackage{color,soulutf8}
\usepackage{graphicx}% Include figure files
\usepackage{dcolumn}% Align table columns on decimal point
\usepackage{bm}% bold math
\begin{document}

%\preprint{APS/123-QED}

\title{Ongoing effects of pandemic-imposed learning disruption on student attitudes}

\author{Teemu Hynninen}
\email{teemu.hynninen@utu.fi}
\author{Henna Pesonen}
\author{Olli Lintu}
\author{Petriina Paturi}
\affiliation{%
Department of Physics and Astronomy, University of Turku, FI-20014, Finland
}%

\date{\today}% It is always \today, today,

\begin{abstract}
We present a study on the development of Finnish first-year physics majors' attitudes towards physics, as measured by the Colorado Learning Attitudes about Science Survey, before, during and after the period of mandatory remote learning due to the coronavirus pandemic. We find that in the years before the pandemic, these attitudes did not change, but the period of extended remote learning due to the pandemic had a negative effect on the students' expert-like views. Similarly, the students who experienced the remote learning period in high school displayed a lower level of expert-like thinking as they entered university. As contact teaching resumed, moderate positive gains were seen, bridging some but not all of the gap in student attitudes left by the pandemic.
\end{abstract}


\maketitle

\section{Introduction}

Many societies shut down in 2020 due to the corona\-virus pandemic. In many countries, this included the complete suspension of contact teaching.
Online courses and distance education were known to be socially demanding with high dropout rates already before the pandemic \cite{rolfe_systematic_2015, Simpson2013, Greenland2022} and the sudden shift to exclusively remote learning is believed to have had and continue to have a profound negative impact on social interactions, engagement and wellbeing of students, both due to the effect it had on the quality of teaching and the social environment in which learning took place \cite{Elmer2020,Rudenstine2021,marzoli_effects_2021}. It is also yet unclear how far reaching in the future the impacts of this disruption will be.

While covid has had a negative impact on multiple aspects in the lives of students, results on how emergency distance education has affected to their academic achievement are somewhat controversial: according to some studies, university students have performed better in distance education caused by the pandemic than they did before the pandemic \cite{burkholder_absence_2022,Iglesias-Pradas2021,Gonzalez2020,Alam2021}. It seems that this positive shift is related to a change in students' learning habits. I.e., in remote teaching, studying is done more continuously rather than just before the exam. On the other hand, many studies imply that emergency distance education has had a negative impact on student performance \cite{Elsalem2021,Giusti2021,Tan2021}. It seems that some student groups, like students at the beginning of their studies or with low self-organization and communication skills, with anxiety or depressive symptomatology, or student's that are lacking academic social interactions are the most vulnerable to the negative effects of remote learning \cite{klein_studying_2021, Giusti2021, Ivanec2022, Tang2021}. 

The impacts of the coronavirus pandemic have been studied from many viewpoints including students' habits \cite{Galle2020}, wellbeing \cite{Rudenstine2021, elmer_students_2020, Giusti2021}, integration \cite{Resch2022}, satisfaction to teaching \cite{Su2021} and academic performance \cite{Iglesias-Pradas2021, klein_studying_2021, kuhfeld_projecting_2020}. However, research on epistemological impacts is almost absent. Only few studies have looked into the changes in students' attitudes towards physics during the pandemic \cite{fox_lab_2021, borish_undergraduate_2022}. Much of the covid-related research is also focusing on younger students, while higher education level is getting less attention.

While student wellbeing and the development of their understanding are important, we are more interested in how extended remote learning has affected the attitudes and views of the students towards physics as a science and themselves as physicists. In our view, the central goal of undergraduate studies in physics is to guide the students' personal growth towards the mindset of a physicist, and this may arguably be better represented by their views instead of academic merits. We study these views using the Colorado Learning Attitudes about Science Survey (CLASS) \cite{adams_new_2006}, a widely used and thoroughly verified tool with which instructors can evaluate how closely the views of their students resemble those of experts, and how teaching affects the development of these views. The views measured by CLASS may even predict the likelihood that a student goes on to become a physicist \cite{perkins_who_2010}.

We administer the CLASS survey yearly to new physics students at the University of Turku, Finland, and study the development of student views during the first year of studies.
Changing attitudes is often a slow process affected by many factors besides teaching, and development towards the mindset of a professional physicist will often continue even after graduation.
Therefore it is realistic to expect only modest development of views during the students' first year of studies. Furthermore, not only is it difficult to develop expert-like thinking, poor physics teaching often has a negative effect on the students' views on physics \cite{madsen_how_2015,Zhang2017,Shi2020}, and the pandemic may have increased the risk for that to happen.

In this work, we set out to generally investigate how the views and attitudes of new physics students have evolved during their first year of studies at the university level before, during and after the pandemic. We are specifically interested in comparing the development of expert-like thinking in three student populations:
\begin{itemize}
\item the students who began their studies in 2019 or earlier and were able to complete their first year of studies in normal conditions,
\item the students who began their studies in 2020 and had to study remotely during their first year of studies, and
\item the students who began their studies in 2021 and had to study remotely in high school before starting their studies at the university.
\end{itemize}
We wish to find out if these groups have developed differently and if so, how.
By comparing the first two groups, we get to compare the effects of contact and remote teaching directly. By comparing the third group to the first two, it is possible to investigate the long term effects a previous teaching disruption has on future learning even after contact teaching has become available again. This kind of information may prove important in the immediate future, as new generations of students who have lived through the pandemic begin their studies.

\section{Course structure}

Some 40--60 physics students begin their studies each year, following a fairly fixed curriculum during their first year of studies.
They study the basics of mechanics, thermodynamics, electromagnetism and wave dynamics during a two-semester calculus-based course series called \emph{Introduction to physics}. 
These courses were originally redesigned and built in 2015-2017 based on ideas such as team-based learning \cite{michaelsen_team-based_2002, johnson_educational_2009}, flipped learning \cite{crouch_peer_2001} and gamification \cite{sailer_gamification_2020}.
The goal was to foster a learning environment that supports active learning and student engagement, promoting both learning and inclusion in the campus community. The courses were not designed to specifically develop expert-like thinking.

The \emph{Introduction} courses are worth about 1/3 of the credits the students earn during their first study year. The rest of the curriculum consists of introduction to studies, tutorials on basic computer use, experimental labs, foreign languages and mathematics courses.

\subsection{Learning before the pandemic}

Before the pandemic, there were no restrictions on contact teaching. We will call this the \emph{full contact teaching} period even though the courses had online components also during this time.

Students were distributed in small teams at the start of each \emph{Introduction} course. These teams were assigned by the teacher and they were constructed to be as heterogeneous as possible. All contact teaching was set to make students actively work with their team. Evaluation was designed so that a student could earn the top grade without taking an exam. Instead, it was enough to participate actively and show enough effort and mastery in course assignments. This was also the preferred way for the students, as only very few students ended up taking the exam.

As the courses started, students first engaged with new information in reading assignments, where the material was distributed online on the social learning platform \emph{Perusall} \cite{miller_use_2018}. This was followed by contact teaching where student teams worked on questions and tutorials, the teacher acting as an advisor.  

Once students became more familiar with the most important physical concepts, the teams were given a few fairly broad and sometimes open-ended problems. The solutions to these problems were turned in on paper at a later date, and given to other teams for peer assessment. While the problems were solved in teams, each problem was also accompanied by questions about the solution process, and every student had to answer these individually.

Traditional physics problems focusing on relatively simple calculations with little or no real physical context were not a major part of these courses. There were examples of such problems in the reading material, and some were also solved as part of the activities during contact teaching, but much more focus was given on conceptual understanding. Still, we provided the students with an online collection of automatically assessed physics problems to complement the concept-focused materal. Solving these problems was not mandatory but could earn credit towards higher grades.

\subsection{Learning during lockdown}

As contact teaching became impossible during the semesters of fall 2020 and spring 2021, all activity had to be taken online. We will call this the time of \emph{no contact teaching}.

At this point we knew that the challenges of keeping students engaged and fostering a feeling of inclusion would become even more crucial than before. However, since the courses were originally designed to promote exactly these factors, we decided that the basic structure of the courses should remain unchanged. 
This meant that students were still assigned to teams, and they still began their studies by reading the given material online. The questions, tutorials and problems were mostly the same or similar to those given to students during normal contact teaching, only now they were distributed online. The same online collection of extra problems was made available.

The courses also followed the same schedule as before lockdown. Whenever there would have been contact teaching in normal conditions, the students would now log on to the online platform \emph{Discord} \cite{discord_inc_discord_2020}, where they would discuss the problems in teams with access to the teacher for advice. If a tutorial would have been completed during the contact teaching session, the students would now submit their work online at the end of the remote study session. The few teacher-driven sessions were replaced by online videos with embedded questions for the teams to answer. Peer assessment was also organized online.

The most important difference compared to previous years was therefore not in what the students were supposed to do but how and where they would do it. Most importantly, all team work and social activities would happen remotely instead of in person.

\subsection{Learning after lockdown}

During the semesters of fall 2021 and spring 2022, the pandemic was ongoing but most people had already access to vaccines. This allowed the university to start a gradual return to normalcy. Contact teaching was allowed in small study groups, but larger gatherings of several tens of people were still banned. This ban was lifted mid semester in the fall, but reinstated at the turn of the year for several more months. In practice, this meant that for half of the academic year, roughly 1/3 of teaching could be given in person, as before the pandemic, while 2/3 was still online. We call this \emph{limited contact teaching}.

\section{Methodology}

We have administered the Finnish translation of the CLASS survey \cite{class_finnish} to our students as part of the \emph{Introduction} courses for four consecutive years, starting in fall 2018. The students take the survey as a pre-survey on the first lesson of the course, at the very beginning of their studies. They take the survey again the next spring near the end of the spring semester as a post-survey. Normally, the survey is given on paper during a lesson to the students present at the time. During remote learning, the survey was made available online and course credit was given for completing the survey in order to ensure a high response rate, as recommended \cite{madsen_administering_2020}. In our analysis, we only consider the students who completed both surveys.

The survey consists of 42 statements answered on a 5-point Likert scale (strongly disagree -- strongly agree). The results are analyzed by calculating expert-like (favorable) and non-expert-like (unfavorable) responses in the pre and post-surveys for each student. These results are compiled in 8 categories (\emph{real world connection}, \emph{personal interest}, \emph{sense making / effort}, \emph{conceptual connections}, \emph{applied conceptual understanding}, \emph{problem solving general}, \emph{problem solving confidence} and \emph{problem solving sophistication}) and an overall score for expert-like thinking. Scores are calculated for the whole class of students by averaging the scores of individual students. \cite{adams_new_2006}

The difference, or effect size, between the score averages of two surveys, $x_0$ and $x_1$, are evaluated using Cohen's $d$,
\begin{equation}
d = \frac{x_1 - x_0}{\sqrt{\frac{1}{2}(s_1^2+s_0^2)}},
\end{equation}
where $s_0^2$ and $s_1^2$ are the corresponding sample variances. A large $d$ means a more pronounced difference between the means of the two distributions.
Although there is no absolute scale for $d$, a common rule of thumb says that $d = 0.5$ corresponds to moderate difference.

\begin{table}[b]
\caption{\label{Ns}
Number of responses each academic year.}
\begin{ruledtabular}
\begin{tabular}{ccc}
semesters & contact teaching &N \\
\hline
fall 2018, spring 2019 & full & 38 \\
fall 2019, spring 2020 & full & 32 \\
fall 2020, spring 2021 & none & 30 \\
fall 2021, spring 2022 & limited & 40 
\end{tabular}
\end{ruledtabular}
\end{table}

Although the \emph{Introduction} courses are meant for first year physics students, some students retake the courses later during their studies. Also a few students majoring in other natural sciences or mathematics decide to enroll on these courses instead of the somewhat easier physics courses meant for non-physicists. In this study, we wish to concentrate on physics majors who are beginning their studies, and therefore we remove responses from other students from the data. This leaves us with a total of 140 valid responses. The breakdown of respones per academic year is shown in Table~\ref{Ns}. This table also lists the limitations on contact teaching for each year. Because we have data from two years of full contact teaching, academic years 2018--2019 and 2019--2020, we combine the survey results from these years in our analysis (\emph{full contact teaching}, $N = 70$).

\section{Results}

\subsection{Effects of teaching on expert-like thinking}

% Overall CLASS
%
% 2018 pre, favorable: 0.73 +- 0.04
% 2018 pre, unfavorable: 0.13 +- 0.03
% 2018 post, favorable: 0.75 +- 0.04
% 2018 post, unfavorable: 0.11 +- 0.03
%
% 2019 pre, favorable: 0.75 +- 0.04
% 2019 pre, unfavorable: 0.11 +- 0.03
% 2019 post, favorable: 0.76 +- 0.04
% 2019 post, unfavorable: 0.12 +- 0.03
%
% 2018+19 pre, favorable: 0.74 +- 0.04
% 2018+19 pre, unfavorable: 0.12 +- 0.03
% 2018+19 post, favorable: 0.75 +- 0.04
% 2018+19 post, unfavorable: 0.11 +- 0.03
%
% 2020 pre, favorable: 0.71 +- 0.04
% 2020 pre, unfavorable: 0.10 +- 0.02
% 2020 post, favorable: 0.68 +- 0.04
% 2020 post, unfavorable: 0.12 +- 0.02
%
% 2021 pre, favorable: 0.69 +- 0.04
% 2021 pre, unfavorable: 0.11 +- 0.03
% 2021 post, favorable: 0.71 +- 0.04
% 2021 post, unfavorable: 0.13 +- 0.03

\begin{figure}[tb]
\includegraphics[width=\columnwidth]{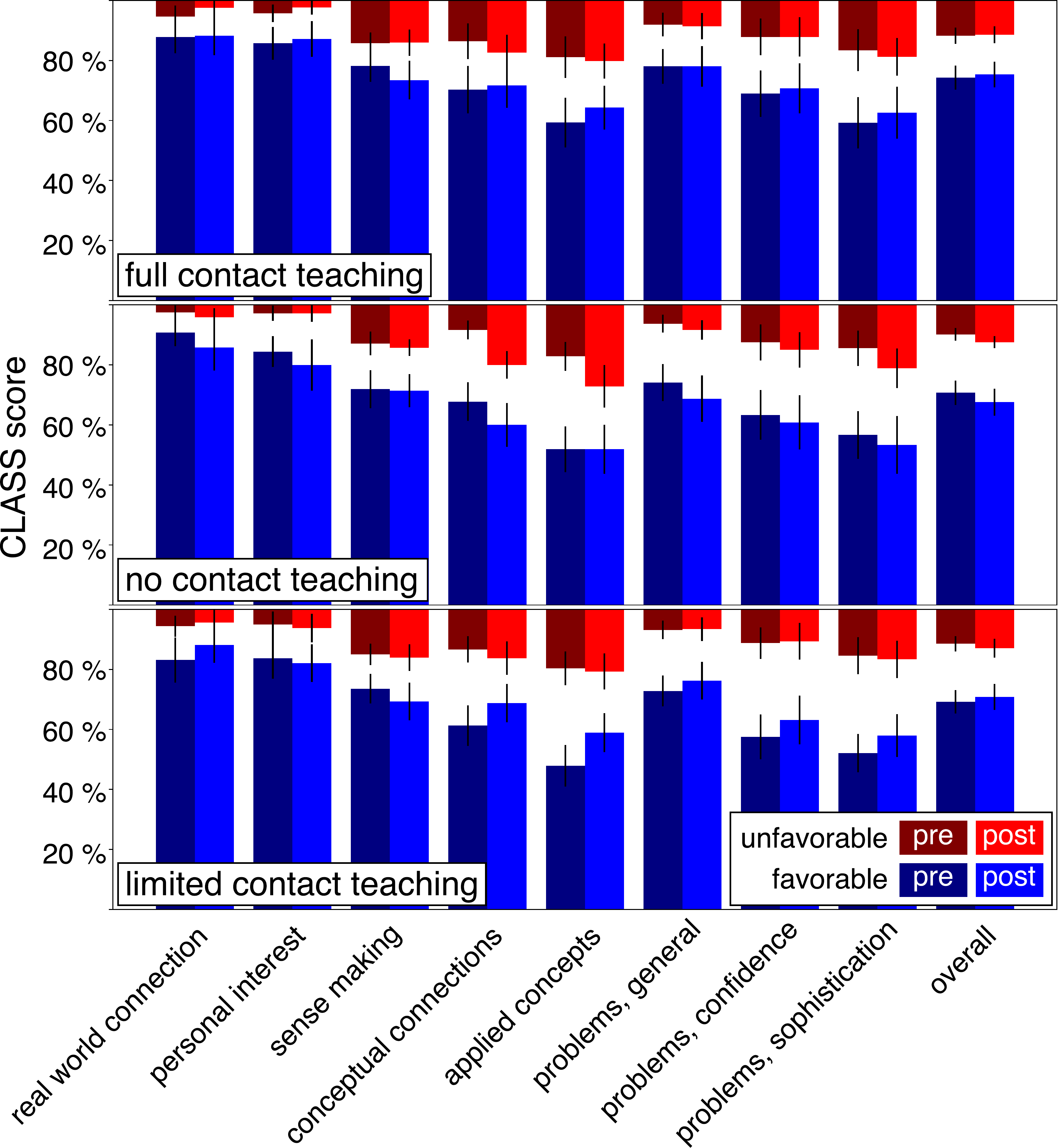}
\caption{\label{category_combo} Portion of expert-like (favorable) and non-expert-like (unfavorable) answers in both pre and post-survey for different levels of contact teaching.}
\end{figure}

Survey results are collected in Fig.~\ref{category_combo} for all three levels of contact teaching. Results from full contact teaching is shown on the top, no contact teaching in the middle, and limited contact teaching at the bottom. Each pair of columns represents a category of the CLASS survey, the rightmost one being the overall score. In each pair of columns, pre-survey results are on the left and post-survey on the right. Bottom bars show the percentage of answers displaying expert-like thinking, and top bars show non-expert-like thinking. The gap in between is the portion of neutral answers. Error bars show the calculated standard error of mean.

Overall, the students display a fairly high level of expert-like thinking. This is no surprise, as we are surveying physics majors, who very likely have a higher than average level of interest in physics. This is seen especially in the categories \emph{real world connection} and \emph{personal interest}, where scores of over 80 \% favorable answers are reached. The lowest scores are seen in the categories \emph{applied conceptual understanding} and \emph{problem solving sophistication}, which measure the students attitudes when applying their knowledge of physics. These categories do not measure just interest but also strategies, which are often still unsophisticated when students begin their studies at the university level.

The effect of teaching is seen by comparing the pre and post-survey results. During the prepandemic years where full contact teaching was possible, shifts in expert-like thinking are fairly moderate. Some gains are made in the categories where the pre-survey scores were the lowest, but there is practically no gain in the overall score. Still, this is not a bad result, since the overall score is fairly high in both pre and post-survey ($74\ \% \pm 4\ \%$ and $75\ \% \pm 4\ \%$ favorable, respectively).

The results are very different when no contact teaching was possible, as the portion of favorable answers are seen to decrease while unfavorable answers become more common. Although in some categories the shifts are not very pronounced, what is most striking is that this happens consistently. Not a single category shows any positive gains. 
Fortunately, when limited contact teaching was again possible, we again see increases in favorable answers, and in some cases these gains seem to be even stronger than when full contact teaching was possible.

\begin{figure}[tb]
\includegraphics[width=\columnwidth]{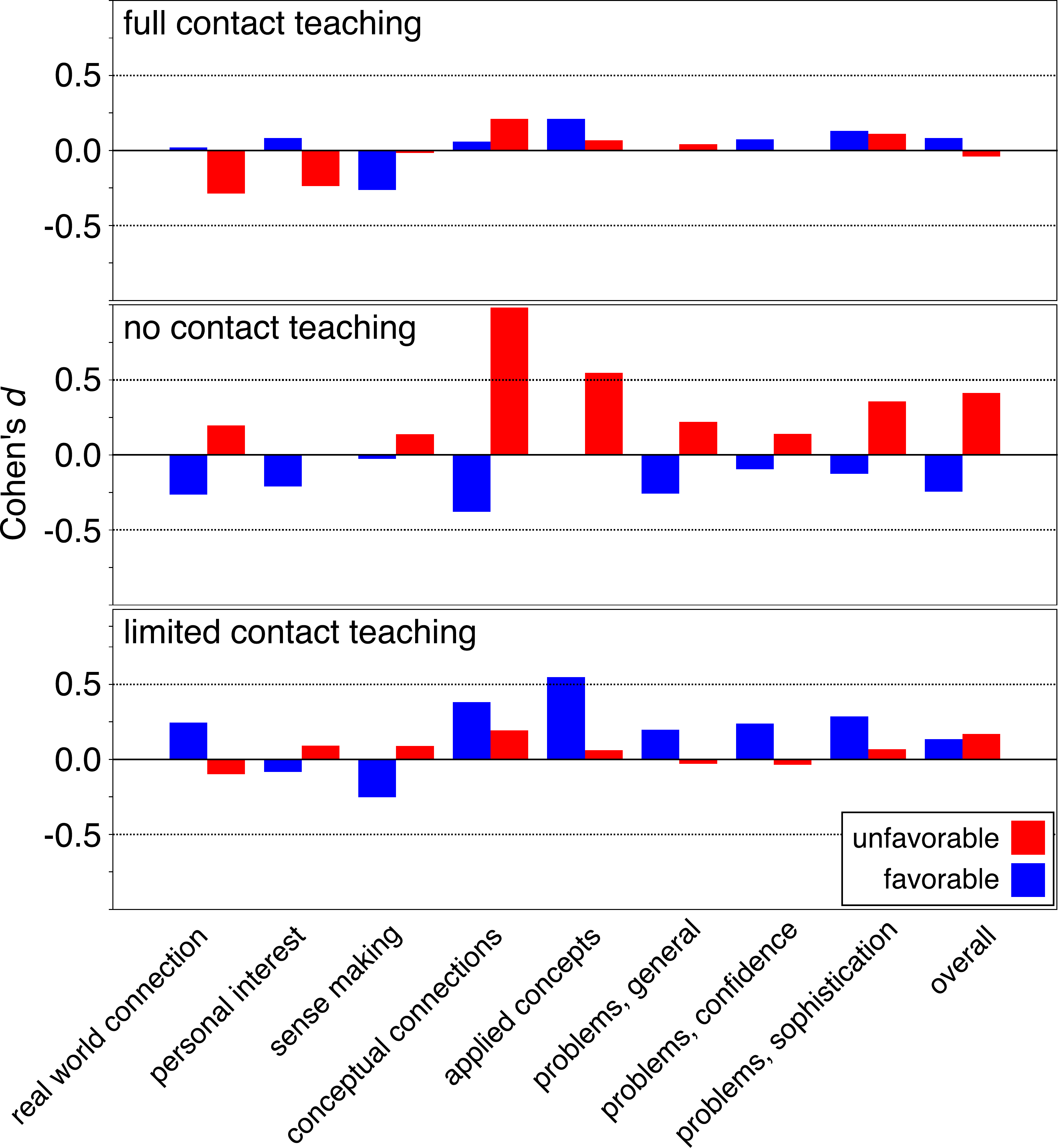}
\caption{\label{cohen_combo} Effect size from pre-survey to post-survey for different levels of contact teaching.}
\end{figure}

The changes from pre to post-survey can be better understood by examining the effect sizes as shown in Fig.~\ref{cohen_combo}. In this plot, a positive value means a higher value in the post-survey. Therefore we wish to see positive values for favorable answers (more expert-like answers after teaching) and negative values for unfavorable answers (less non-expert-like answers after teaching).

As we already discussed, full contact teaching seems to have had little effect on student thinking. We see positive and negative values for both favorable and unfavorable answers, but the effect sizes are typically close to zero and at most $|d| = 0.26$, which is still considered to be a small effect.

The case is different for the students who received no contact teaching. In all categories, the effect on favorable answers is either zero or negative while the effect on negative answers is zero or positive. This negative effect is small or moderate on the favorable answers, but in some categories there are strong effects on the unfavorable answers. The strongest increase in unfavorable answers is seen for \emph{conceptual connections} where the effect size is 1.0. Upon further analysis we find that although there is an increase in unfavorable answers to all the statements in this category, this effect is most strongly driven by a great increase of negative answers in statement 1, ''A significant problem in learning physics is being able to memorize all the information I need to know.'' In the pre-survey, only 10 \% of students agreed with this statement while 40 \% did so in the post-survey. For comparison, 19 \% and 17 \% of students in full contact teaching agreed with this statement in pre and post-surveys, respectively. 

Besides statement 1, unfavorable answers increase markedly also in statement 5, ''After I study a topic in physics and feel that I understand it, I have difficulty solving problems on the same topic'' (from 35 \% to 57 \%), and 40, ''If I get stuck on a physics problem, there is no chance I’ll figure it out on my own'' (from 20 \% to 30 \%).
These contribute to the \emph{applied conceptual understanding} as well the \emph{problem solving} categories, and as a result these categories also show a moderate effect in increased non-expert-like thinking.

Much more positive results are seen with limited contact teaching. Although overall gains in favorable answers are small and there is roughly equal increase in unfavorable answers, most categories show small to moderate effects in increased favorable answers. In fact, the highest gains in expert-like thinking are seen in \emph{conceptual connections} and \emph{applied conceptual understanding}, where we saw the highest increase in non-expert-like thinking for the students with no contact teaching. All \emph{problem solving} categories also show a small positive effect in expert-like thinking. These results can again be attributed to statement 1, 5 and 40, only now we see a significant increase in favorable answers (from 43 \%, 8 \% and 25 \% to 58 \%, 18 \% and 48 \%, respectively).

\subsection{Distributions of expert-like thinking scores}

\begin{figure}[tb]
\includegraphics[width=\columnwidth]{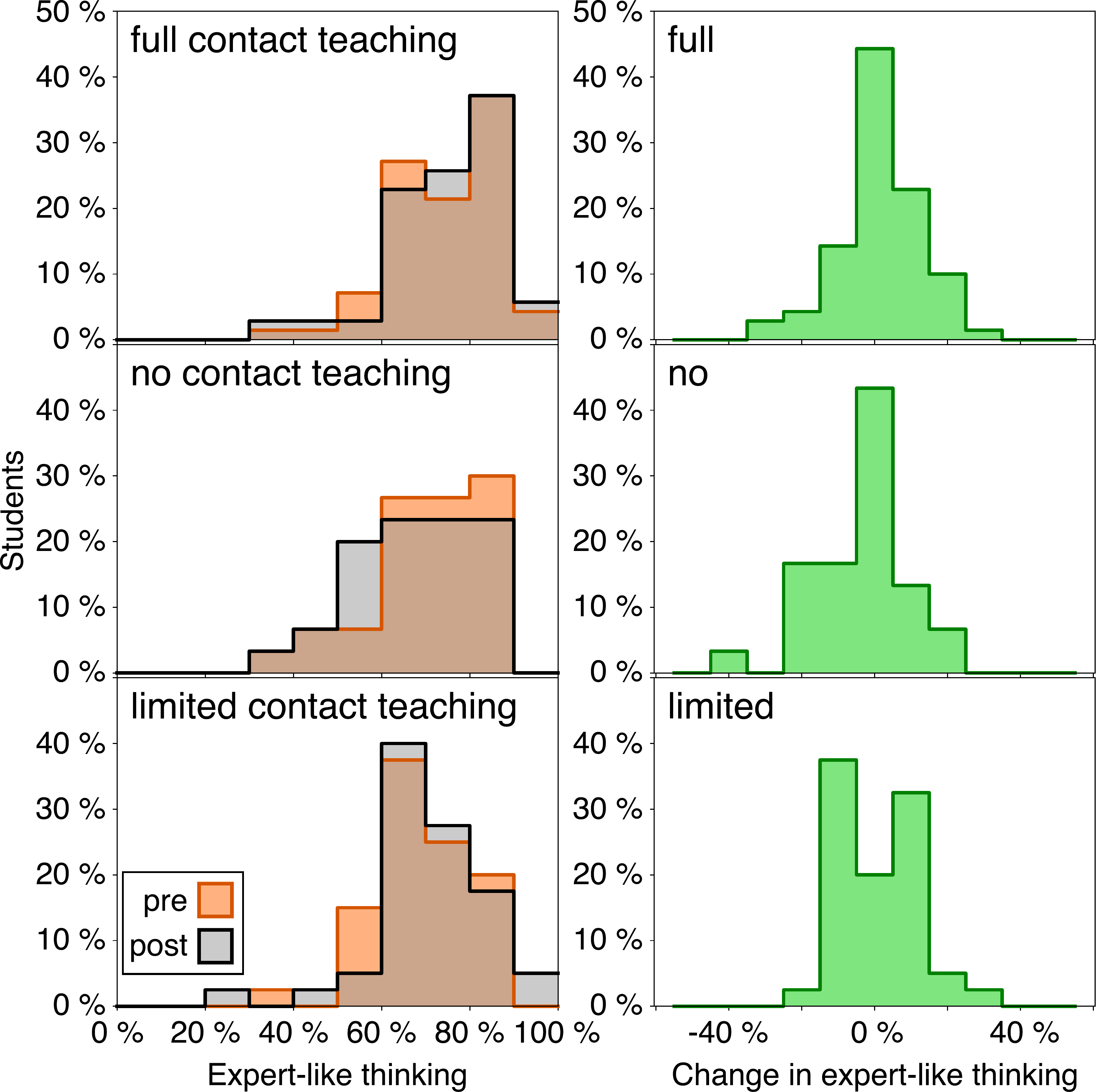}
\caption{\label{histograms} Distribution of overall favorable CLASS scores in pre and post-surveys, and the change in this score.}
\end{figure}

Next, we turn our attention from mere averages to the distribution of expert-like thinking in the student population. Fig.~\ref{histograms} show these distributions for overall favorable answers in both pre and post-surveys, as well as the distribution of change in favorable answers.

In classes that received full contact teaching, most students demonstrate expert-like thinking in 60 \% to 90 \% of their answers in both pre and post-surveys, as only a few students fall outside of this range. The distribution peaks in the 80 \% to 90 \% range showing that many students have developed a high level of expert-like thinking already in high school. The distribution for the change in expert-like thinking peaks symmetrically at 0 \% showing that, on average, student views did not change.

The class that had to study with no contact teaching shows a similar distribution in the pre-survey with most students in the 60 \% to 90 \% range. However, this is not true anymore for the post-survey distribution, which has broadened so that the bulk of students now fall in the 50 \% to 90 \% range. That is, the number of students scoring between 60 \% and 90 \% has decreased while those scoring between 50 \% and 60 \% has increased. This drift towards lower scores in expert-like thinking is seen also in the distribution for change, which again peaks at 0 \% but is now asymmetric, displaying more negative than positive changes.

The situation is reversed in the class that studied under limited contact teaching. Now, the bulk of students are distributed between 50 \% and 90 \% already in the pre-survey. Furthermore, the distribution peaks between 60 \% and 70 \%. Clearly, this student population demonstrated a lower level of expert-like thinking compared to previous years at the start of their studies.

In the post-survey most students are again found in the 60 \% to 90 \% range with only few students in the 50 \% to 60 \% interval, showing a positive shift due to teaching. However, the distribution still peaks between 60 \% to 70 \% instead of 80 \% to 90 \% as in the full contact teaching data. Also, the distribution for changes in expert-like thinking now shows two peaks at around $\pm 10 \ \%$. Curiously, further analysis reveals this double peak to be entirely due to female students. However, we could not demonstrate any particular explaining factor for the split in the distribution.

\begin{figure}[tb]
\includegraphics[width=\columnwidth]{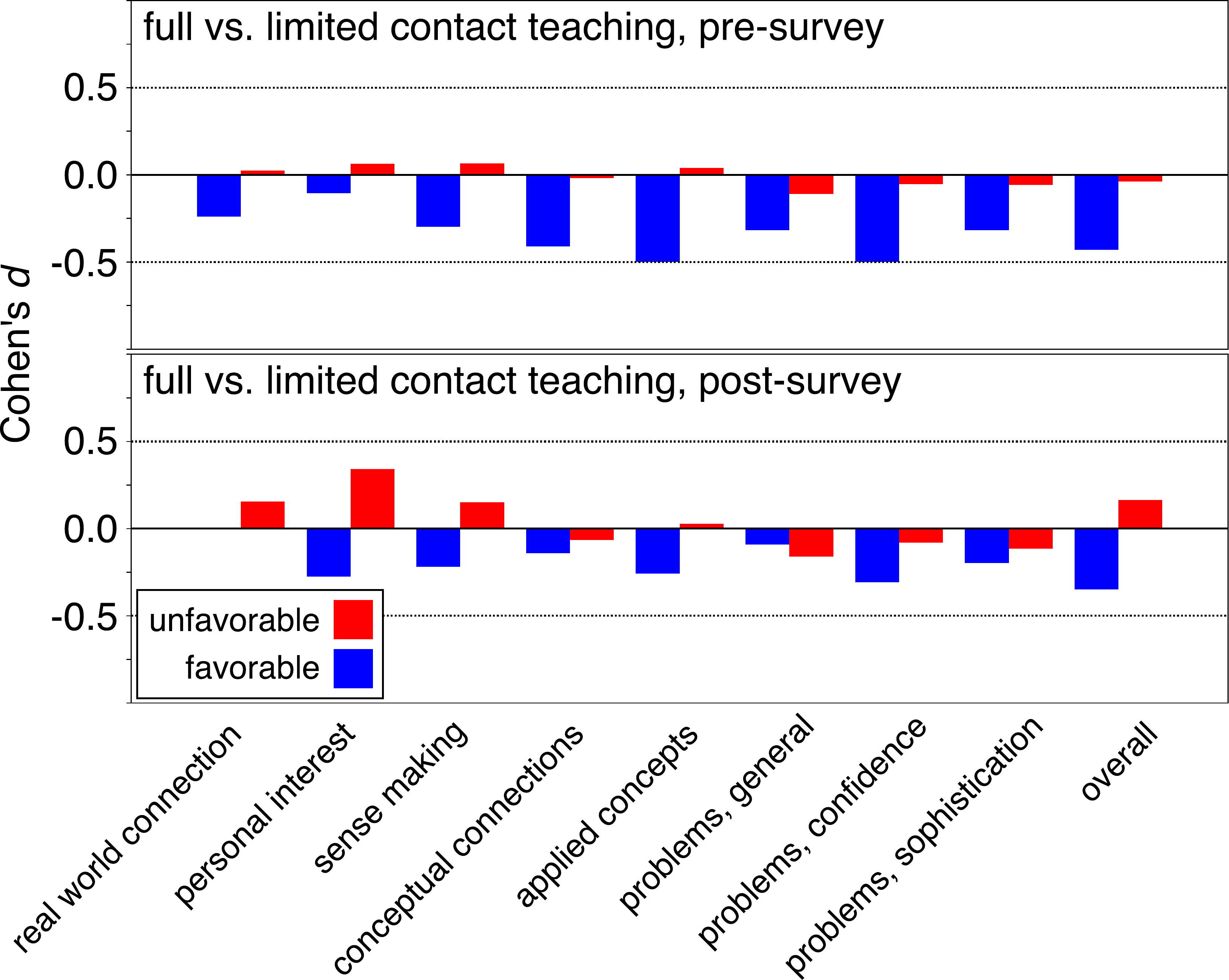}
\caption{\label{normal_to_limited} Comparison of survey results for classes with full and limited contact teaching. Differences are calculated in such a way that a positive value means the class with limited contact teaching scored higher.}
\end{figure}

To further quantify the differences between the full and limited contact teaching classes, we calculate Cohen's $d$ for their pre and post-survey results in all CLASS categories, as shown in Fig.~\ref{normal_to_limited}. Note the difference to Fig.~\ref{cohen_combo}: We are not comparing the same set of students before and after teaching, so this is not a measure of the effect of teaching. We are measuring statistical difference between distributions involving different students. The plot on the top shows the difference in pre-survey between full and limited contact teaching groups, and the bottom plot similarly for post-survey. Positive values mean a higher score was recorded for the limited contact teaching group.

Comparing favorable answers in pre-surveys, we find that the group that received limited contact teaching scored lower in all categories compared to the group in full contact teaching. In several categories, including the overall CLASS score, $|d| \approx 0.5$, meaning that the difference is moderately significant. Furthermore, even though we saw generally positive gains in expert-like thinking during limited contact teaching, these gains were not enough to close the gap, as the limited contact teaching group still scored lower in the post-survey compared to the the full contact teaching group.

\subsection{Course completion}

% R:
% typically R < 0.35
% no contact:
% personal interest post-survey R = 0.54
% connections post-survey R = 0.46
% applied concepts R = 0.49 -> 0.46
% problem general R = 0.52 -> 0.51
% problem confidence R = 0.62 -> 0.55
% problem sophistication R = 0.51 -> 0.54
% overall R = 0.45 -> 0.65

Our survey data has shown that there are moderate differences in the level of expert-like thinking in students who studied under different conditions. Finally, we compare these results to student performance.

Since students pass the \emph{Introduction} courses by participating in course activities, grades do not reflect mastery of the subject, as displayed effort is a fairly significant factor in several assignments.
Still, the final grade is a measure of some combination of motivation, effort and mastery, and therefore a meaningful reference nonetheless.

To test if there is correlation between CLASS results and course performance, we calculate Pearson's correlation coefficient $R$ between the CLASS category scores and the final grade (given on an integer scale of 0--5).
As a result, we do find positive correlation, but usually this correlation is fairly weak. For classes in full or limited contact teaching, all $R$-values are about 0.35 or less. So even though high-performing students tend to have higher CLASS scores on average, CLASS is not a good predictor of individual student performance in these groups.

However, for students that received no contact teaching, we see stronger correlation between CLASS and grades. The highest $R$ is found for post-survey overall CLASS score, for which $R = 0.65$. There is moderately strong correlation also between all the \emph{problem solving} categories and the final grade, as they have $R$-values between 0.5 and 0.6 in both pre and post-survey ($p < 0.001$ for all).

\begin{table}[tbh]
\caption{\label{pass_rate}
Number of physics majors who passed the \emph{Introduction} courses in the fall and in the spring and the dropout rate.}
\begin{ruledtabular}
\begin{tabular}{cccc}
semesters & passed & passed & dropout \\
 & fall & spring & rate \\
\hline
fall 2018, spring 2019 & 56 & 52 & 7 \% \\ % grade: $3.8 \pm 0.2$ 
fall 2019, spring 2020 & 44 & 40 & 9 \% \\ %  $3.8 \pm 0.3$
fall 2020, spring 2021 & 45 & 37 & 17 \% \\ % $3.6 \pm 0.2$
fall 2021, spring 2022 & 52 & 49 & 6 \% % $3.0 \pm 0.3$
\end{tabular}
\end{ruledtabular}
\end{table}

Finally, in Table~\ref{pass_rate}, we list the total number of physics majors who passed the \emph{Introduction} courses each fall and spring, and the corresponding dropout rate calculated as the difference between these two numbers, divided by the number of students who pass the fall course.
In the years 2018--2019 and 2019--2020, with full contact teaching, the dropout rate during the first study year was slightly less than 10 \%.
In 2020--2021 and no contact teaching, the dropout rate doubled to 17 \%. This is in accordance with negative changes in expert-like thinking during this year.
Lastly in 2021--2022 and limited contact teaching, a low dropout rate of 6 \% was recovered.

\section{Discussion}

We set out to compare the development of expert-like views in three different groups of new physics majors, and our results found differences between all of them.

Before the pandemic, students displayed a fairly high level of expert-like thinking already at the beginning of their studies, and their views did not significantly change during their first year of studies. The students who began their studies in 2020 also demonstrated similarly high levels of expert-like views at the beginning of their studies, but their results deteriorated during a year of remote learning, showing a systematic decrease in expert-like answers and an increase in non-expert-like answers. This result is not surprising but it is in direct contrast to previous results that showed no change in the students' views on experimental physics after emergency remote teaching \cite{fox_lab_2021}. Similarly in a study conducted before the pandemic, there were no differences in attitudes between students in in-person and online physics laboratory teaching \cite{Rosen2020}. 

Although student views developed negatively in all categories, the most severe effect was on a statement about physics as a collection of facts to remember. There is lot to learn during undergraduate studies, but physics forms a logical whole which must be studied through understanding instead of memorization. A student who sees physics as a collection of facts is not in a position to construct a coherent understanding of physical theories and will likely face severe difficulties in his or her studies. Therefore, it is a worrying observation that the pandemic affected this view so negatively. This result is also in agreement with some earlier studies where emergency remote teaching had a negative impact on student performance \cite{Elsalem2021,Giusti2021,Tan2021}. 

The flipped learning model applied in the \emph{Introduction} courses makes the student responsible for getting acquainted with new information, but the logical structure of this information is explored together with the teacher during lessons. In contact teaching, this was succesful, but even though students had access to teacher assistance during the pandemic, many students failed to understand these connections when this aid was provided remotely. This is possibly due to it being easier for the students to miss or even purposefully avoid communicating with the teacher and their peers in an online environment, while it is more difficult for the teacher to reach the students in order to guide them. In such an environment, self-organization and communication skills of the students become increasingly important \cite{klein_studying_2021}. Also, understanding the logical structure of physics requires resilience which many students may have lacked in the stressful environment created by the pandemic \cite{balta-salvador_academic_2021,Pertegal-Felices2022}.

We also saw that the overall CLASS score as well as scores in the \emph{problem solving} categories were fairly good indicators of academic performance in the no contact teacing group, unlike in the other groups. This suggests that a contact teaching environment can help overcome some of the non-expert-like views students have regarding problem solving. Perhaps this is because students seek help more frequently face-to-face than online \cite{Mahasneh2012,Reeves2015}. In a remote learning environment, students benefit much more from adopting an expert-like approach to problem solving. The necessity for well-developed study skills is likely reflected also in the dropout rate \cite{Robbins2004,Bas2018,Polansky1993}, which was clearly higher in remote learning compared to other groups. It seems that contact learning could provide enough aid and social support so that also most of the low-performing students were able to complete the courses, whereas remote learning could not.

The students in the limited contact teaching group began their studies a year later after having experienced the emergency remote learning during their final year in high school. Subsequently, they scored consistently lower in expert-like thinking in the pre-survey compared to previous classes. We cannot claim for certain nor confirm that this lower score is due to the learning disruption they experienced earlier, as we have no CLASS data for this group from before the pandemic. This difference can be due to other factors or even just chance. Also, a few students in this group had graduated from high school earlier than the previous spring and had not experienced the pandemic during high school. Therefore it is an oversimplification to claim that the drop in expert-like thinking is solely due to remote teaching. It is plausible, however, that earlier remote teaching is was an important contributing factor.

In any case, as we saw in the group that received no contact teaching, a disruption in the learning environment can cause deterioration in expert-like thinking. Therefore we may consider the group in limited contact teaching as a \emph{model} for a population whose views have been affected by the pandemic, even if we are not certain if that the pandemic was the ultimate reason for the observed difference.

For this group of students, participation in the flipped-learning-based \emph{Introduction} courses resulted in an overall increase in both expert-like and non-expert-like thinking. Importantly though, this group displayed positive development in expert-like thinking in exactly the same statements that were most adversely affected by the pandemic in the previous class. The dropout risk was also reduced back to pre-pandemic level despite possible skill-deficits compared to earlier classes.
This suggests that negative effects of earlier learning disruptions could be gradually mended once contact learning can be resumed. 
Still, overall non-expert-like thinking increased as much as expert-like thinking in this group, and their post-survey CLASS score was still moderately lower than that of the full contact learning group. Apparently, the negative effects from one year of disrupted studies can be long-lasting and not easily mended in a similar timeframe.

On a more positive note, these results were obtained in limited contact teaching. This indicates that even though many students do not cope well in a learning environment based only on remote learning, providing some regular in-person lessons in addition to online activities may be enough to mitigate the negative effects of exclusive remote learning.

\bibliography{PER}% Produces the bibliography via BibTeX.

\end{document}